**Revealing Wavelength- and Size-Dependent $CO_2$ Reduction Selectivity via Operando Scanning Photo-Electrochemical Microscopy**

*Fatemeh Kiani[1], Milad Sabzehparvar[1], Priscila Vensaus[1], Elif Nur Dayi[1], Olga D'Anania[2,3], Tarique Anwar[1], Nuria Lopez[2], Ravishankar Sundararaman[3,4], Giulia Tagliabue[1]**

*[1] Laboratory of Nanoscience for Energy Technologies (LNET), STI, École Polytechnique Fédérale de Lausanne, 1015 Lausanne, Switzerland*

*[2] Institute of Chemical Research of Catalonia (ICIQ-CERCA), The Barcelona Institute of Science and Technology, Av. Països Catalans, 16, 43007 Tarragona, Spain*

*[3] Scuola Superiore Meridionale, Largo San Marcellino, 80138 Naples, Italy*

*[4] Department of Materials Science & Engineering, Rensselaer Polytechnic Institute, 110 8th Street, Troy, New York 12180, USA*

**E-mail: giulia.tagliabue@epfl.ch*

**Abstract**

Controlling product selectivity in plasmonic catalysis, particularly in $CO_2$ reduction ($CO_2R$), remains a central unsolved challenge with direct implications for light-driven fuel and chemical synthesis. Here, we deploy quantitative operando scanning photoelectrochemical microscopy (photo-SECM) to provide a direct demonstration that tuning photon energy switches $CO_2R$ selectivity through an electronically driven pathway. On plasmonic Au/p-GaN photocathodes, interband excitation (460-560 nm) drives selective CO production while intraband excitation (640-800 nm) favors $H_2$ evolution. By maintaining constant absorbed power across wavelengths and confirming linear power dependence, we isolate the role of hot-carrier energy from photonic and photothermal contributions. Density functional theory calculations reveal that higher-energy interband excitation progressively increases the overlap between hot-electron-accessible states and the CO-producing intermediate, selectively promoting CO over formate, in excellent agreement with experiment. We further show that selectivity is geometrically gated by hot-carrier transport: sub-100 nm nanostructures sustain $CO_2R$ activity, while ~300 nm nanodisks suffer transport losses that suppress it, consistent with ab initio hot-carrier transport calculations. Together, these results establish photon energy, carrier transport, and nanostructure geometry as coupled design parameters for plasmonic $CO_2R$ selectivity, resolve a longstanding debate on the origin of plasmon-driven selectivity effects, and position photo-SECM as a broadly applicable operando platform for photo(electro)catalysis.

## Introduction

Photo(electro)catalytic $CO_2$ reduction ($CO_2R$) offers a promising route to convert solar energy into carbon-neutral fuels and value-added chemicals. Plasmonic metal nanostructures are unique photocatalysts, as they synergistically combine catalytic activity with the ability to harvest visible light[1,2]. Strong near-field enhancement, hot-carrier (energetic electrons and holes) generation, and local heating create energy-rich conditions that can drive $CO_2R$[3–9], with hot-carrier transfer in particular enabling access to excited states that steer selectivity in multi-pathway $CO_2$ reduction reactions ($CO_2RR$)[4,7–9]. When integrated with wide-bandgap semiconductors, such as p-type GaN, plasmonic metals have emerged as effective photocathodes[4,5,10]. In this architecture, p-GaN favors charge separation by efficiently extracting hot holes above the Schottky barrier, while hot electrons remain in the metal to synergistically lower kinetic barriers for $CO_2R$. Yet, how the energetics and transport of hot electrons, modulated by photon energy and nanostructure size, govern product selectivity under operando conditions remains poorly understood due to the lack of sufficiently sensitive in-situ techniques.

Scanning electrochemical probe microscopy techniques have recently emerged as powerful operando tools to investigate electrocatalytic processes at the nano- and microscale[11,12], outperforming the conventional electrochemical methods such as rotating ring-disk electrodes[13,14]. Among them, scanning electrochemical microscopy (SECM) stands out for its ability to locally detect bulk-reaction products[15], enabling the analysis and mapping of product distributions (selectivity)[16,17] and the quantification of catalytic activity with very high sensitivity (down to tens of nanomolar)[12], different from optical spectroscopic methods which typically probe surface-bound intermediates on the catalyst surface[18,19]. Using a biased ultramicroelectrode (UME) probe, SECM re-oxidizes the $CO_2R$ products generated at the substrate, with UME tip voltammetry allowing electrochemical discrimination of species such as CO, formate, and $H_2$[20]. Unlike bulk analytical methods such as gas chromatography (GC), nuclear magnetic resonance (NMR) spectroscopy, and mass spectrometry, which provide accurate product quantification but lack spatiotemporal resolution and are typically applied ex-situ, SECM enables fast, sensitive and in-situ detection of $CO_2R$ products at the site and moment of their formation[20–22]. These advantages have enabled SECM to be applied for detecting short-lived rection intermediates such as $CO_2^{\cdot-}$ radical anion (with a half-life of ~10 ns)[23], to resolve the main origin of formate formation in bicarbonate electrolytes[20], to elucidate CO oxidation mechanisms as a function of interfacial pH[24], to map facet-dependent selectivity on Au surfaces[25], and to screen catalyst arrays with compositional or morphological variations[21,22]. This capability makes SECM particularly valuable for high-throughput catalyst screening and mechanistic investigations prior to more time- and resource-intensive bulk analyses. However, while GC and NMR remain the standard techniques for quantitative product analysis, SECM has never been benchmarked against them, and such validation is essential to establish SECM as both a qualitative and quantitative tool for local and real-time probing $CO_2R$. Furthermore, SECM applications have so far been limited to purely electrochemical conditions, and the influence of light excitation particularly wavelength-dependent hot-carrier processes in plasmonic photocathodes on $CO_2R$ activity and selectivity has remained unexplored. Extending SECM into the photoelectrochemical domain therefore provides a unique opportunity to uncover how different light conditions (photon energy and flux) and nanostructure geometry govern product distribution under operando conditions.

In this work, we demonstrate that scanning photo-electrochemical microscopy (photo-SECM, also known as SPECM) is a quantitative operando platform for photochemistry and leverage it, for the first time, to quantify the wavelength- and size-dependent selectivity of $CO_2RR$ on plasmonic Au/p-GaN photocathodes in situ. By combining spectral tuning of incident photons with local product detection via UME voltammetry, we resolve distinct transitions in product selectivity between the gold interband and intraband excitation regimes. We show that, at constant absorbed light power, interband excitation (460-560 nm) promotes $CO_2RR$ while suppressing the competing $H_2$ evolution reaction (HER), with CO increasingly favored at higher photon energies, whereas intraband excitation (640-800 nm) favors $H_2$ and suppresses CO production. Density functional theory calculations rationalizes this result by revealing that O*CO intermediate, the precursor for CO production, has a distinctive high-energy density of states that can be populated by hot-electrons generated by photons at energies higher than ~2eV, consistent with experimental observations. Furthermore, comparison across different Au nanostructures reveals a pronounced size effect. Ab-initio carrier transport calculation demonstrates that smaller gold nanotriangles (<100 nm) exhibit high and spatially uniform hot-carrier surface fluxes, sustaining efficient $CO_2RR$, while larger gold nanodisks (≃300 nm) suffer from transport losses that suppress $CO_2RR$ and shift selectivity toward HER. Overall, these results establish an electronically driven pathway for controlling plasmonic $CO_2RR$ selectivity through photon energy, while highlighting the coupled roles of hot-carrier energetics, carrier transport, and nanostructure geometry. More broadly, they position photo-SECM as a powerful operando platform for uncovering structure-mechanism-selectivity relationships in photo(electro)catalysis. Understanding and modeling these mechanisms are crucial for the rational design of highly efficient wavelength- and size-tailored plasmonic catalysts.

## Results and Discussion

We fabricated plasmonic photocathodes consisting of arrays of gold nanotriangles (Au NTs, lateral size ≃ 70 nm), gold nanodisks (Au NDs, diameter ≃ 300 nm), and dispersed gold nanoparticles (Au NPs, average size of ≃ 34 ± 7 nm) supported on an optically transparent p-type GaN (p-GaN) wide-bandgap semiconductor (bandgap ≃ 3.35 eV, **Figure S5**). SEM images of the fabricated structures are shown in

**Figure 1a**. The Au NTs and NDs were prepared using nanosphere lithography[26,27], while the Au NPs were obtained by a sputtering-annealing technique (see Methods in **SI 1.1**, **Figures S1-2**). We optically characterized the samples[28] to determine their absorption spectra (see Methods in **SI 1.2**, **Figure S4**). Au NTs and NDs structures show a broad dipole plasmon resonance mode in the intraband region (750-800 nm), which enables us to disentangle plasmon absorption and interband excitation effects. We studied $CO_2$RR on these plasmonic photocathodes in a $CO_2$-saturated 0.1 M cesium bicarbonate ($CsHCO_3$) electrolyte using photo-SECM, as schematically illustrated in **Figure 1a** (see Methods in **SI 1.3**). We performed photo-SECM measurements in substrate generation/tip collection (SG/TC) mode, where the substrate was biased at −1.5 V vs Ag/AgCl cathodic potential and illuminated from the bottom with a collimated laser beam (≃ 30μm diameter), while recording cyclic voltammograms (CVs) at a Pt UME tip. The potential of −1.5 V was chosen as an optimal condition that provided sufficient product generation and stable SECM signals without inducing vigorous bubble formation or compromising the stability of the GaN substrate. Applying higher cathodic potentials leads to the formation of oxynitride phases on GaN[29], which enhances its intrinsic catalytic activity and convolutes the observed Au-driven catalytic behavior.

We positioned the Pt UME tip 10 μm (≃10 times the Pt core radius) above the illuminated substrate area, where the photoinduced reactions occur. The products of bicarbonate and $CO_2$ reduction (formate ($HCOO^-$) and CO), along with $H_2$ from the competing hydrogen evolution reaction (HER), desorb from the substrate and diffuse toward the Pt UME, and after adsorbing onto its surface, are subsequently removed by oxidation under the applied potential at the tip. In this process, CO and formate were oxidized back to $CO_2$, and $H_2$ was oxidized to water.

To benchmark the measurements, we compared the voltametric responses of the substrate and the UME tip. The substrate CV (**Figure 1b**) displayed only a monotonic rise in cathodic current with increasing negative potential under illumination, with no resolved product features. In contrast, the UME CV (**Figure 1c**) showed distinct peaks corresponding to the formate oxidation reaction (FOR), the CO oxidation reaction (COOR), and the hydrogen oxidation reaction (HOR), consistent with previous electrochemical studies[20,21,24]. Characteristic signatures of CO and formate oxidation appeared in the forward and backward scans, respectively, while a broad bump between −0.3 and −0.6 V was attributed to $H_2$ oxidation in the forward scan. The area under each feature can then be used to quantitatively estimate the amount of products generated by the substrate. To confirming the ability of SECM to quantitatively resolve product selectivity in real time, we compared the SECM results to standard bulk product quantification using gas chromatography (GC) and proton nuclear magnetic resonance ($^1$H-NMR) spectroscopy. Specifically, we performed electrochemical $CO_2$R on a planar Au foil at different cathodic potentials using SECM, while in parallel employing an independent electrochemical setup coupled to GC for gas product analysis (CO and $H_2$) from the same sample.

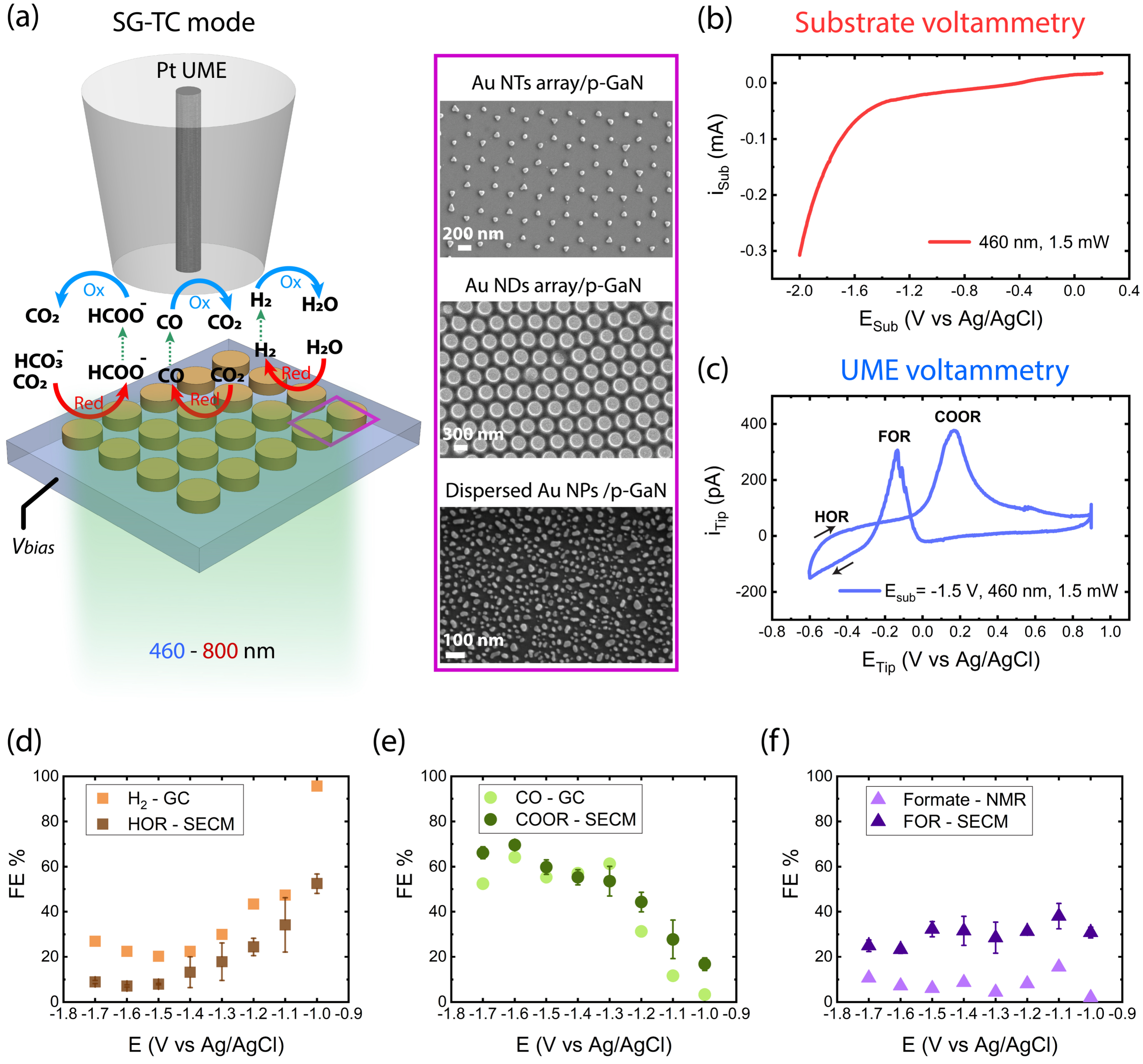

**Figure 1.** *Benchmarking photo-SECM UME tip voltammetry against substrate voltammetry and bulk analytical methods.* (a) Schematic of the substrate generation/tip collection (SG/TC) mode in photo-SECM, showing reduction of $CO_2/HCO_3^-$ at the Au/p-GaN substrate and oxidation of products at the Pt UME tip in a $CO_2$-saturated 0.1 M $CsHCO_3$ electrolyte. SEM images of Au nanotriangle (NT) and nanodisk (ND) arrays, and dispersed nanoparticles (NPs) on p-GaN are shown on the right. (b) Substrate CV at –1.5 V vs Ag/AgCl under 460 nm illumination, showing a monotonic rise in cathodic current without resolved product features. (c) Corresponding UME CV, showing distinct oxidation peaks assigned to the formate oxidation reaction (FOR), CO oxidation reaction (COOR), and hydrogen oxidation reaction (HOR). (d-f) Comparison of Faradaic efficiencies (FEs) obtained by SECM at the UME tip with bulk product quantification by gas chromatography (GC, for CO and $H_2$) and $^1$H-NMR (for formate) during electrochemical $CO_2$RR on a planar Au foil at different applied cathodic potentials.

Liquid products (formate) were quantified separately by $^1$H-NMR (see Methods in **SI 1.4-5, SI 3**, and **Figures S6-9**). The results (**Figures 1d-f**) show that Faradaic efficiencies (FEs) determined by SECM are comparable in magnitude and trends to those obtained from GC and $^1$H-NMR across most applied potentials (particularly from –1.1 V to more negative values), despite being computed based on the UME tip current rather than the substrate current. Observed deviations can be explained by the local nature of SECM measurements compared to the bulk-averaged GC and $^1$H-NMR analyses, which obscure spatial variations in $CO_2$R products (**Figure S6**). Remarkably, SECM exhibits higher sensitivity toward formate detection, likely due to its ability to probe products directly and in real time at the reaction site, whereas $^1$H-NMR relies on sampling hundreds of microliters of bulk electrolyte, which can dilute or attenuate product signals. Consistent with this, the total FE obtained from GC and $^1$H-NMR is slightly below 100% (**Figure S9**), reflecting the limited sensitivity of $^1$H-NMR toward formate quantification. To the best of our knowledge, such a direct comparison of SECM with bulk quantitative methods has not been reported previously. These results demonstrate that SECM can provide not only qualitative detection of multiple $CO_2$RR products but also quantitative analysis with high sensitivity, positioning it as a powerful operando tool for mechanistic studies and catalyst screening.

Having established this, we first focused on Au NT photocathodes as a representative system to study the effect of photon energy and irradiation power on $CO_2$R product formation. **Figure 2a** shows representative CVs recorded at the UME tip under illumination with 460 nm and varying incident powers (50-1500 µW). Distinct oxidation peaks corresponding to FOR, COOR, and HOR are clearly observed, with their relative intensities depending on the incident power. CVs recorded at other wavelengths and for the Au NDs and Au NPs are provided in **Figures S10, S12**, and **S14**, respectively. Control experiments under the same conditions on bare p-GaN substrates, in the absence of Au nanostructures, showed no activity or voltammetric features associated with product oxidation (**Figures S11** and **S13**), confirming that the observed reduction reactions originate solely from hot-carrier generation in Au. To further compare the sample response at different wavelengths while excluding possible variations arising from wavelength-dependent absorption and thermal effects, it is critical to account for the absorption spectrum of the sample. **Figure 2b** shows representative CVs recorded at the UME tip under illumination at different excitation wavelengths but equal absorbed power (460-800 nm, absorbed power = 0.3 mW). Clearly, $CO_2$R product distribution exhibits a pronounced wavelength-dependence, indicating that photon energy directly influences selectivity.

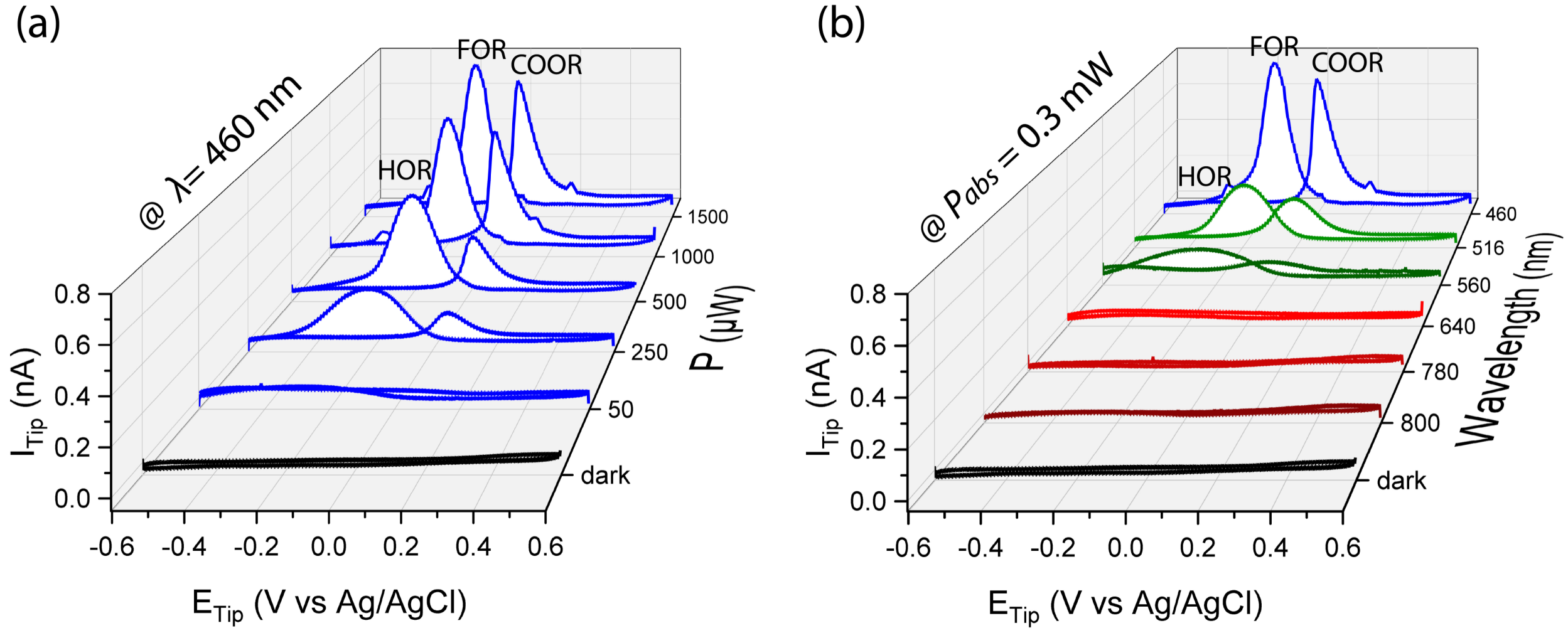

**Figure 2.** *Wavelength- and power-dependent UME tip voltammetry for Au NT/p-GaN photocathodes.* (a) Representative CVs recorded at a 1 μm-radius Pt UME tip at dark and under illumination (a) at 460 nm excitation with varying incident powers (50-1500 μW), and (b) at different excitation wavelengths (460-800 nm) at an equal absorbed power of 0.3 mW . Distinct oxidation features corresponding to FOR, COOR, and HOR are observed, with their relative intensities depending on photon energy and incident power.

From the tip currents in the CVs (**Figure 2b**), we quantified the consumed charge (Q) and Faradaic efficiencies (FE) for product oxidation at the UME tip for AuNT/p-GaN photocathode (**Figures 3a** and **3b**, eq. S1 in **SI 1.3**). The results reveal a distinct selectivity transition between the interband (d-sp) and intraband (sp-sp) excitation regimes of Au. At shorter wavelengths (i.e. higher photon energies) in the interband region (460-640 nm), CO and formate production is strongly enhanced. Notably, CO formation occurs exclusively under interband excitation, with FE reaching a maximum of ≃ 40% at 460 nm. In contrast, at longer wavelengths (i.e. lower photon energies) in the intraband region (640-800 nm), $H_2$ formation dominates, with its FE increasing up to ≃ 60% while CO formation vanishes. Formate displays intermediate behavior, remaining nearly constant across the intraband regime but peaking sharply at 560 nm (80% FE) near the interband threshold. These trends clearly demonstrate that interband excitation favors $CO_2$RR to CO and formate while suppressing HER, whereas intraband excitation shifts selectivity toward $H_2$.

For a given nanostructure and excitation wavelength, the prompt hot carrier energy distribution is fixed. When we varied the incident power, the $Q_{product}$ data (**Figure S15**) increased linearly with absorbed power, consistent with a hot-carrier generation rate scaling linearly with photon absorption[30,31]. In contrast, a thermally driven process would be expected to show an exponential dependence on power[32,33]. Moreover, the Faradaic efficiencies remained nearly constant with changes in absorbed power at a given wavelength (**Figure S16**), in contrast to the significant selectivity shifts observed across different wavelengths (**Figure 3b**). These observations indicate that the reactions are governed by hot-carrier-driven processes, with negligible contributions from photothermal effects.

The remaining question is how the observed $CO_2$RR selectivity switch between interband and intraband regimes is related to hot-carrier energetics as well as any transport and charge separation effects. **Figure 3c** schematically illustrates the prompt hot carrier energy distribution in Au at five different excitation wavelengths and the energy diagram of the Au/p-GaN system. Hot electrons with energies $\geq$ 1 eV are all capable, in principle, of driving the reduction reactions leading to $H_2$, CO and formate ($HCOO^-$) formation. Yet, the selectivity towards different $CO_2$R processes is related to the adsorption configuration of $CO_2$ molecule at the catalyst surface[34,35]. To this end we employed DFT to understand photocatalytic product distribution. We modelled different surfaces, representing Au(111), open Au(100), and defect-like structures Au(211); details are provided in section **SI 1.6** of the Methods. When $CO_2$ approaches the surface, a physisorbed (initial) state appears where $CO_2$ keeps its linearity, this was taken as the reference ground state in the simulations. We then considered two potential adsorbed states: the $\eta_{1,C}$ O*CO where both C and O are connected to the surface and serves as the precursor to CO, and the $\eta_{2,O,C}$ *OCO* intermediate that accounts for formate production. These two states are metastable and highly endergonic and can therefore only be populated once the intial state has been excited (i.e. when electrons are promoted to energy regions accessible through interband or intraband excitation). By computing the overlap[36,37] between the aligned density of states (DOS) of the two intermediates and that of the linear $CO_2$ configuration, we estimate the probability of energy transfer, which correlates with the corresponding product formation rates (see **SI 1.6** for details). **Figure S18** shows the overlapped density of states for the $CO_2$ configurations on Au(111), illustrating the distribution of available electronic states as a function of energy. We note that the energy axis is aligned to the Au Fermi level illustrated in **Figure 3c** to allow for direct comparison with the hot-carrier energetics. The percentage overlap in the DOS areas between each intermediate and that of linearly adsorbed $CO_2$ was calculated according to eq. S9 and plotted as a function of energy in **Figure 3d**. Focusing on the energy ranges corresponding to the excitation wavelengths, we observe that at lower energies (up to around 1.9 eV), the contributions of the two configurations are similar for Au(100) and Au(211), with a dominance of *OCO* for Au(111) (**Figures 3d**, **S19**, and **S20**). At higher energies, the contribution of the OC*O intermediate becomes more pronounced for all investigated facets, indicating enhanced CO production. From a comparison with the prompt hot-carrier energy distributions in **Figure 3c**, it is clear that this intermediate high energy state can be populated by hot electrons only when using photon energies above the interband threshold, with a major increase for photon energies above 2.2 eV (~560 nm wavelength), in excellent agreement with the experimental observations.

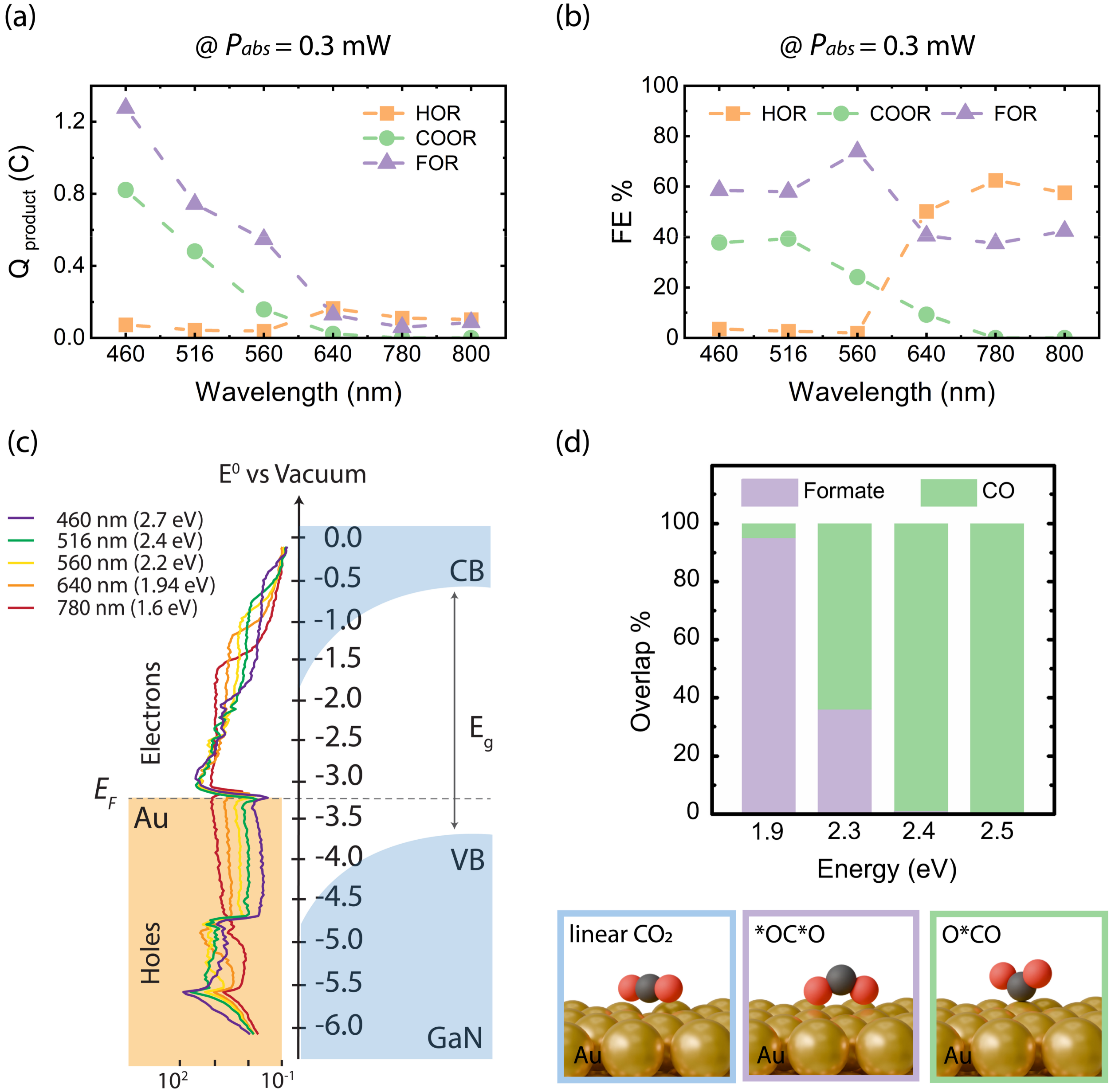

**Figure 3.** *Wavelength-dependent $CO_2$RR product selectivity in photo-SECM on Au NTs/p-GaN photocathodes.* (a) Consumed charge (Q) for CO, formate, and $H_2$ oxidation at the Pt UME tip under illumination at different excitation wavelengths at a fixed absorbed power of 0.3 mW. Higher photon energies (interband regime, 460-560 nm) enhance CO and formate production, while lower photon energies (intraband regime, 640-800 nm) favor $H_2$. (b) Faradaic efficiencies (FEs) for COOR, FOR, and HOR at the Pt UME tip as a function of wavelength, revealing a selectivity switch between interband and intraband excitation: CO FE maximizes at 460 nm, $H_2$ dominates at longer wavelengths, and formate peaks at 560 nm. (c) Calculated hot-carrier energy distributions in Au under excitation at different wavelengths (logarithmic flux scale) and the corresponding energy-band alignment of the Au/p-GaN interface at −1.5 V

vs Ag/AgCl. (d) Percentage overlap between the DOS of linearly adsorbed $CO_2$ (reference state, blue) and the intermediates associated with formate (*OC*O, lavender) and CO (O*CO, green) formation on Au(111) as a function of excitation energy. The overlap was calculated within ±0.1 eV of each excitation energy. Corresponding optimized structures are shown below.

Yet, carrier transport also influences the energy distribution of hot carriers available at the Au surface. We thus calculated the cumulative carrier fluxes reaching the surfaces of the nanostructures directly or upon scattering under illumination in both interband and intraband regimes (see Methods in **SI 1.7** and **Figures S21** and **S22**). At shorter wavelengths (e.g. 460 nm, 2.7 eV), interband excitation generates an initial hot carrier distribution with a peak of high-energy holes and lower-energy electrons, together with a small fraction of higher-energy electrons reaching energies close to the photon energy. In our previous work[10], we showed that hole transfer at the Au/p-GaN interface is most efficient under interband excitation, due to the presence of high-energy d-band holes capable of overcoming the approximately 1.3 eV Schottky barrier[10]. At 460 nm, this results in maximum charge separation, reducing recombination and increasing higher-energy electrons that can be injected into carbon-adsorbed $CO_2$ molecules, favoring CO production. In contrast, at longer wavelengths, intraband transitions yield a nearly uniform carrier distribution up to the photon energy, reducing hole extraction efficiency, as well as the population of high-energy electrons. As a result, recombination increases and the availability of sufficiently energetic electrons to populate the O*CO intermediate decreases, suppressing CO formation (**Figures 3a** and **b**). Finally we note that the consistently higher production of formate compared to CO across all excitation wavelengths is due to the presence of direct bicarbonate reduction, as demonstrated experimentally (Supplementary Information 6, **Figure S23**). Indeed, although we kept the electrolyte under continuous $CO_2$ flow to stabilize the bulk pH during the reaction, local alkalization inevitably occurs during the reduction reactions due to hydroxide ion generation, favoring bicarbonate-mediated pathways towards formate formation[20]. Overall, in our system the formation of CO is the key indicator of the selectivity switch. Modelling results clearly show that this is driven by the injection of high energy electrons into O*CO, boosted by improved charge separation at the Au/GaN interface, both significantly enhanced at photon energies >2eV due to the inter-band transitions in Au.

The interplay of adsorbate energetics and carrier transport can be further analyzed by comparing the three studied plasmonic structures, which reveal notable differences in product distribution under identical excitation conditions. For the smaller structures, such as Au NTs and dispersed NPs, the wavelength-dependent FE trends are very similar (**Figure 4a** and **Figure S17**). In contrast, NDs, which are significantly larger, exhibit distinct product distributions and FE trends (**Figure 4b**). For NDs, formate formation peaks and CO production occurs only at the highest photon energy (2.7 eV, 460 nm), where hot-carrier energies are sufficient to overcome transport losses. At photon energies below 2.7 eV, enhanced carrier scattering in the larger NDs leads to a sharp decline in product formation, for example, $Q_{COOR}$ decreases from 0.9 C to 0 C, and $Q_{FOR}$ drops from 2 C to ~0.095 C (**Figure 4b**, $Q_{product}$). Interestingly, despite lower surface coverage, Au NTs display trends similar to dispersed Au NPs: CO and formate dominate in the interband regime, while $H_2$ becomes predominant in the intraband regime (**Figure 4a** and **Figure S17**). This behavior can be attributed to their comparable particle sizes (< 100 nm, NT thickness ≃ 10 nm), which facilitate hot-electron transport with fewer scattering events. Higher-energy carriers in Au have a very short life-time and limited mean free path (≃10 nm for 2eV hot electrons and ≃2 nm for 2eV hot holes)[31], which recombine on ultrafast time scales (~ 100s fs) if they are not collected. In contrast, NDs, although having surface coverage similar to Au NPs, are more than three times larger (disk diameter ≃ 300 nm) which results in more transport losses and limits the availability of high-energy electrons for $CO_2$RR. These results suggest that particle size, rather than surface coverage, primarily dictates carrier transport behavior[15,38] and, in turn, $CO_2$RR product distribution across photon energies. The distinct, film-like behavior of NDs, favoring $H_2$ production even at high photon energies in the interband regime, supports this interpretation (**Figure 4b**, FE%).

Previous electrochemical studies have also demonstrated strong particle-size effects on $CO_2$RR selectivity. On Au and Ag, smaller nanoparticles with higher proportions of low-coordination edge/step/kink sites favor CO formation, while larger, more extended surfaces show diminished CO selectivity and increased HER contributions[39,40]. Similar size-dependent selectivity shifts have been reported for Cu, Bi, and $In_2O_3$, often attributed to differences in facet exposure, local electric fields, pH environments, or surface reconstructions[41–43]. Our findings extend these insights to plasmonic hot-carrier-driven catalysis, where the dominant size-dependence arises not only from surface chemistry but also from hot-carrier transport within the nanostructure.

This conclusion is further supported by theoretical transport simulations (**Figure 4c**), which show 3D spatially resolved surface fluxes of hot electrons with energies greater than a threshold ($E_e \geq 1$ eV and 2 eV) incident on the Au NT and Au ND surfaces under two excitation wavelengths corresponding to the interband (460 nm) and intraband (800 nm) regimes. The 1 eV energy threshold corresponds to the energy range containing the majority of hot carriers, whereas the 2 eV threshold showcases the presence of high-energy hot electrons capable of accessing the O*CO unique adsorbate state. These are generated only by interband transitions, while their surface flux is critically affected by their short mean free path. A clear difference in both the magnitude and spatial distribution of hot-electron fluxes is observed between the Au ND and Au NT structures. The maximum flux of hot electrons reaching the Au interfaces is significantly higher for both energy thresholds and under both excitation regimes for the Au NT compared to the Au ND. Moreover, the flux over the Au NT is more uniformly distributed across the surface (also nanotriangles are

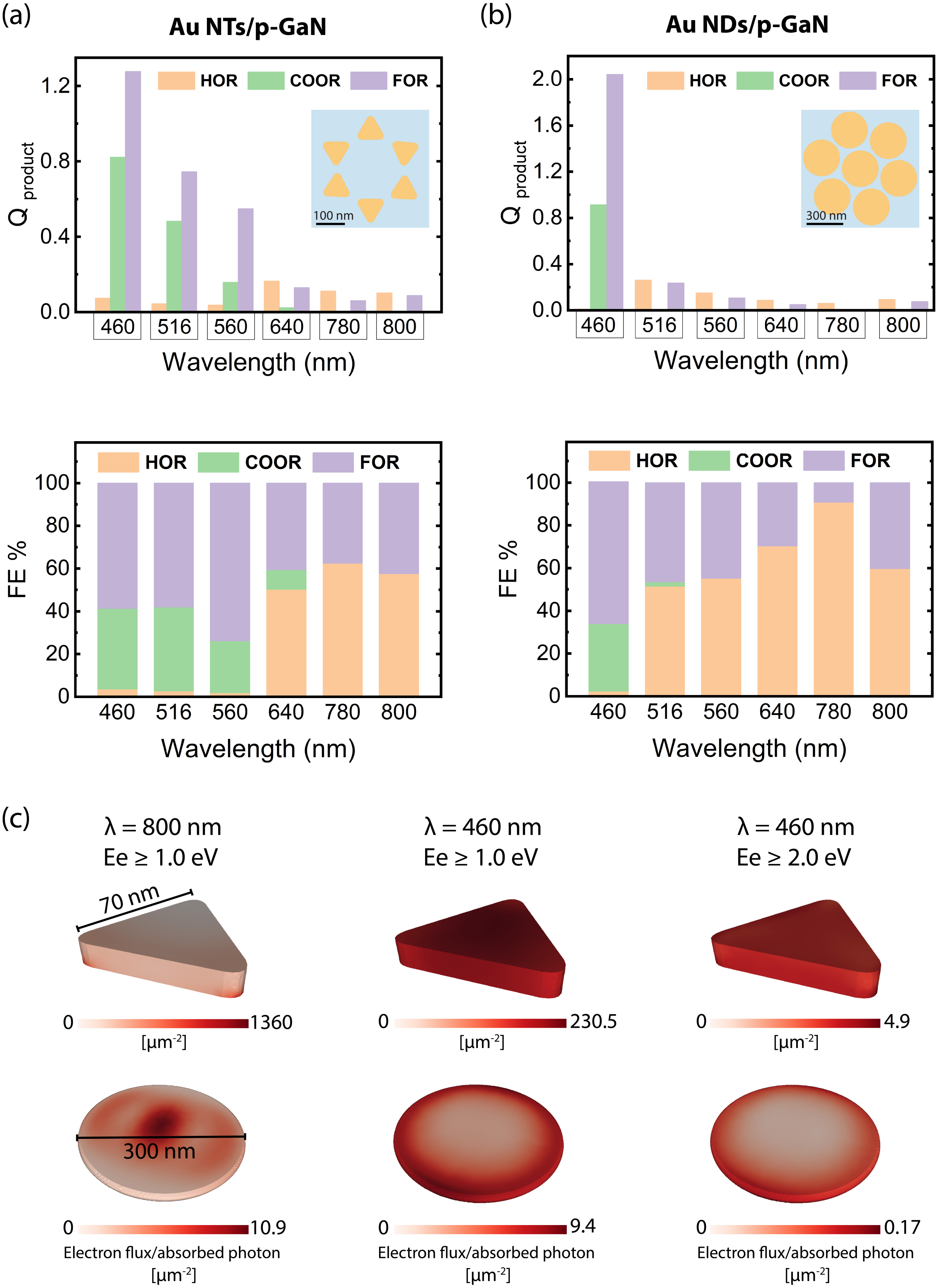
(a)
Au NTs/p-GaN
HOR
COOR
FOR
Q product
100 nm
Wavelength (nm)
(b)
Au NDs/p-GaN
300 nm
FE %
(c)
λ = 800 nm
Ee ≥ 1.0 eV
λ = 460 nm
Ee ≥ 1.0 eV
λ = 460 nm
Ee ≥ 2.0 eV
70 nm
300 nm
[μm-2]
Electron flux/absorbed photon
[μm-2]

**Figure 4.** *Effect of nanostructure size on $CO_2RR$ selectivity and hot-carrier transport in Au/p-GaN photocathodes.* (a, b) Consumed charge ($Q_{product}$, top) and Faradaic efficiencies (FE%, bottom) for COOR, FOR, and HOR at the Pt UME tip under illumination of (a) Au NTs/p-GaN and (b) Au NDs/p-GaN photocathodes at excitation wavelengths of 460 -800 nm (absorbed power = 0.3 mW). Au NTs (≃ 70 nm) show enhanced $CO_2RR$ activity with CO and formate dominating in the interband regime and $H_2$ in the intraband regime, while larger Au NDs (≃ 300 nm) exhibit suppressed $CO_2RR$ and favor $H_2$ production even at shorter wavelengths. (c) Theoretical simulations of hot carrier transport showing 3D spatially resolved surface fluxes of hot electrons ($E_e \geq 1$ eV and 2 eV) for Au NTs and NDs under 460 and 800 nm illumination. Fluxes are normalized to the absorbed photon and unit area of the surface.

likely to expose more Au(111)-like facets), enhancing the probability of transferring hot electrons to the O*CO from both the top and side interfaces exposed to the electrolyte. In contrast, the Au ND (most likely exhibiting more Au(100) domains) exhibits a highly localized flux, concentrated near the edges at shorter wavelengths and toward the center at longer wavelengths. This nonuniformity in surface flux across different interfaces of the ND is also evident from their corresponding calculated energy-resolved carrier fluxes at different surfaces shown in **Figure S22**. The lower and spatially nonuniform flux reduces the probability of injecting hot electrons into O*CO and contributes to the differences in product selectivity observed for NDs. These results highlight the critical role of nanoparticle geometry and size in governing hot-carrier transport and $CO_2RR$ selectivity.

In summary, we proved operando photo-SECM as a quantitative and highly sensitive platform to probe photoelectrochemical processes and we unravelled, for the first time, the microscopic mechanisms of light-driven $CO_2RR$ selectivity steering in plasmonic Au/p-GaN photocathodes. Benchmarking against GC and $^1$H-NMR confirmed that SECM not only resolves product selectivity in real time but also provides Faradaic efficiencies comparable to bulk methods, with enhanced sensitivity toward formate. By tuning the excitation wavelength while maintaining constant absorbed power, we revealed a clear selectivity switch: interband excitation (460-560 nm) generates high-energy hot carriers that are not accessible under intraband illumination and promotes efficient hole extraction across the Au/p-GaN Schottky barrier, enabling charge separation that leaves a larger population of energetic electrons at the Au/electrolyte interface capable to access the highest density of states of O*CO adsorbate, leading to enhanced CO production. In contrast, intraband excitation (640-800 nm) results in lower hole extraction efficiency and fewer high-energy electrons, leading to suppressed $CO_2RR$ and dominant $H_2$ production. Notably, the nearly constant Faradaic efficiencies across different absorbed powers confirm that the observed selectivity trend is hot-carrier-driven rather than photothermal. A pronounced size dependence was also observed: smaller triangular nanostructures (NTs, NPs <100 nm) maintained efficient $CO_2$RR under interband excitation, whereas larger nanodisks (~300 nm) suffered hot-carrier scattering losses, reducing the flux of energetic electrons at the surface and shifting selectivity toward HER even at high photon energies. Consistent with this, our theoretical transport simulations revealed lower and less-uniform surface fluxes of hot-electrons in NDs compared to NTs.

Together, these results establish photo-SECM as a versatile operando tool to uncover the interplay of photon energy, power, and nanostructure geometry in plasmonic photocatalysis, and reveal an electronically driven mechanism for plasmonic-assisted $CO_2$RR selectivity, governed by the interplay between hot-carrier energetics at selected excitation wavelengths and carrier extraction and transport, providing design principles for wavelength- and size-tailored plasmonic catalysts.

**Supporting Information**

Methods section; additional information about characterizaion of the Au/p-GaN heterostructures; benchmarking SECM with GC and $^1$H-NMR; wavelength- and power-dependent UME voltammetry analysis of Au/p-GaN Photocathodes; hot carrier transport prediction in Au/p-GaN heterostructures. All the DFT data has been uploaded in the ioChem-BD repository[44,45] at https://iochem-bd.iciq.es/browse/review-collection/100/120400/91334a59ed389d92c579431b and http://dx.doi.org/10.19061/iochem-bd-1-429.

**Author Information**

Corresponding Author

Giulia Tagliabue − *Laboratory of Nanoscience for Energy Technologies (LNET), STI, Ecole Polytechnique Fédérale de Lausanne, 1015 Lausanne, Switzerland*; orcid.org/0000-0003-4587-728X, Email: giulia.tagliabue@epfl.ch

Authors

Fatemeh Kiani − *Laboratory of Nanoscience for Energy Technologies (LNET), STI, Ecole Polytechnique Fédérale de Lausanne, 1015 Lausanne, Switzerland*; *Laboratory for Multifunctional Materilas, ETH Zürich, 8093 Zürich, Switzerland;* orcid.org/0000-0002-2707-5251

Milad Sabzehparvar − *Laboratory of Nanoscience for Energy Technologies (LNET), STI, Ecole Polytechnique Fédérale de Lausanne, 1015 Lausanne, Switzerland* ; orcid.org/0000-0001-5594-6889

Priscila Vensaus − *Laboratory of Nanoscience for Energy Technologies (LNET), STI, Ecole Polytechnique Fédérale de Lausanne, 1015 Lausanne, Switzerland* ; orcid.org/0000-0002-3859-5578

Elif Nur Dayi − *Laboratory of Nanoscience for Energy Technologies (LNET), STI, École Polytechnique Fédérale de Lausanne, 1015 Lausanne, Switzerland ;* orcid.org/0000-0001-5509-3249

Olga D'Anania − *Institute of Chemical Research of Catalonia (ICIQ-CERCA), The Barcelona Institute of Science and Technology, Tarragona 43007, Spain ; Scuola Superiore Meridionale, Largo San Marcellino, I-80138 Napoli, Italy ;* orcid.org/0000-000-7172-360X
Tarique Anwar − *Laboratory of Nanoscience for Energy Technologies (LNET), STI, École Polytechnique Fédérale de Lausanne, 1015 Lausanne, Switzerland ;* orcid.org/0000-0003-0655-2298
Núria López − *Institute of Chemical Research of Catalonia (ICIQ-CERCA), The Barcelona Institute of Science and Technology, Tarragona 43007, Spain* ; orcid.org/0000- 0001-9150-5941
Ravishankar Sundararaman − *Department of Materials Science & Engineering, Rensselaer Polytechnic Institute, Troy, New York 12180, United States;* orcid.org/0000-0002-0625-4592

**Author Contributions**

F.K. and G.T conceived the project and F.K. led the project implementation. F.K. and M.S. developed the photo-SECM setup, and M.S. fabricated the Pt UME tips. F.K. fabricated all samples, with assistance from T.A. in using the Langmuir-Blodgett technique for nanosphere lithography. F.K. performed all structural and optical characterizations, including absorption spectroscopy, SEM imaging, and Mott-Schottky analysis. F.K. conducted all photo-SECM measurements and data analysis. Gas chromatography (GC) measurements and analysis were performed by F.K. with assistance from P.V., and $^1$H-NMR measurements and analysis were conducted by F.K. and E.N.D. Electromagnetic simulations were performed by F.K., and hot-carrier transport simulations were carried out by R.S. O.D. and N.L. performed the DFT simulations and their analysis. F.K. wrote the manuscript with input from all authors. G.T. supervised all aspects of the project.

**Acknlowledgements**

F.K., M.S., P.V., E.N.D., and G.T. acknowledge support from the Swiss National Science Foundation (Eccellenza Grant No. 194181). M.S. and G.T. also acknowledge the STI Imaging Fund, supported by the EPFL Center for Imaging. The authors further acknowledge the use of experimental facilities at EPFL, including the Center of MicroNanoTechnology (CMi) and the Interdisciplinary Centre for Electron Microscopy (CIME). GaN wafers were provided by the Advanced Semiconductors for Photonics and Electronics (LASPE) group and the EPiX Platform at EPFL. Hot-carrier transport calculations using the NESSE framework were performed at the Center for Computational Innovations at Rensselaer Polytechnic Institute. O.D. and N.L. acknowledge the Barcelona Supercomputing Center (BSC-RES) for generous computational support. The authors thank Dr. Gopal Narmada Naidu for advice on the electromagnetic simulations and Dr. Kiseok Oh for guidance on SECM UME voltammetry during the early stages of the project.